\documentclass[showpacs,pra,aps,superscriptaddress,floatfix,tightenlines,twocolumn]{revtex4}

\usepackage{amsmath}
\usepackage{amssymb}
\usepackage{amsfonts}
\usepackage{bm}
\usepackage{times}
\usepackage{amsthm}
\usepackage{subfigure}
\usepackage{color}
\usepackage{graphicx}

\usepackage[colorlinks,bookmarks=false,citecolor=blue,linkcolor=red,urlcolor=blue]{hyperref}

\begin{document}
\title{Phase diagram of quantum critical system via local convertibility of ground state}
\author{Si-Yuan Liu }
\affiliation{Institute of Modern Physics, Northwest University, Xian
710069, P. R. China }
\affiliation{Beijing National Laboratory for Condensed Matter Physics,
Institute of Physics, Chinese Academy of Sciences, Beijing 100190, P. R. China}

\author{Quan Quan}
\affiliation{Institute of Modern Physics, Northwest University, Xian
710069, P. R. China }
\affiliation{School of Mathematical Sciences, Capital Normal University, Beijing 100048, China}

\author{Jin-Jun Chen}
\affiliation{Beijing National Laboratory for Condensed Matter Physics,
Institute of Physics, Chinese Academy of Sciences, Beijing 100190, P. R. China}

\author{Yu-Ran Zhang}
\email{yrzhang@iphy.ac.cn}
\affiliation{Beijing National Laboratory for Condensed Matter Physics,
Institute of Physics, Chinese Academy of Sciences, Beijing 100190, P. R. China}

\author{Wen-Li Yang }
\email{wlyang@nwu.edu.cn }
\affiliation{Institute of Modern Physics, Northwest University, Xian
710069, P. R. China }

\author{Heng Fan}
\email{hfan@iphy.ac.cn}
\affiliation{Beijing National Laboratory for Condensed Matter Physics,
Institute of Physics, Chinese Academy of Sciences, Beijing 100190, P. R. China}
\affiliation{Collaborative Innovation Center of Quantum Matter, Beijing, P. R. China}
\affiliation{Institute of Modern Physics, Northwest University, Xian
710069, P. R. China }
\date{\today}
\begin{abstract}
We investigate the relationship between two kinds of ground-state local convertibility
and quantum phase transitions in XY model. The local operations and classical
communications (LOCC) convertibility is examined by the majorization relations and the
entanglement-assisted local operations and classical communications (ELOCC) via R\'{e}nyi
entropy interception. In the phase diagram of XY model, LOCC convertibility and ELOCC
convertibility of ground-states are presented and compared. It is shown that different phases in the phase
diagram of XY model can have different LOCC or ELOCC convertibility, which can be used to
detect the quantum phase transition.
This study will enlighten extensive studies of quantum phase transitions from the
perspective of local convertibility, e.g., finite-temperature phase transitions and other
quantum many-body models.
\end{abstract}
\maketitle

\section{Introduction}
Many developments in quantum information processing (QIP) \cite{key-1}
unveiling the rich structure of quantum states and the nature of entanglement
have offered many insights into quantum many-body physics \cite{key-2}.
Because the ground-state wavefunction undergoes qualitative changes at a
quantum phase transition, it is important to understand how its genuine
quantum aspects evolve throughout the transition \cite{key-3}. Many alternative
indictors of quantum phase transitions, for example, entanglements measured by
concurrence \cite{key-3}, negativity \cite{key-4}, geometric entanglement \cite{key-5},
and von Neumann entropy \cite{key-6,key-7} have been investigated in several critical
systems, which have become a focus of attention in detecting
a number of critical points. From the viewpoint
of quantum correlations, mutual information \cite{key-8}, quantum discord \cite{key-9} and
global quantum discord \cite{key-10,gd,key-21} have been used for detecting
quantum phase transitions. There are investigations on entanglement spectra
\cite{key-11,key-12,key-13} and fidelity \cite{key-14} showing their
abilities in exploring numerous phase transition points in various
critical systems as well.

With the concepts of QIP, these studies have achieved great success
in understanding the deep nature of the different phase transitions.
The reverse, however, is still unclear and deserves further investigations. Recently,
from a new point of view, Cui \emph{et al.} find that the systems undergoing
quantum phase transition will also show different operational properties
from the perspective of QIP  both analytically \cite{prx} and numerically \cite{key-15,key-16}. For several models,
they reveal that the entanglement-assisted local operations and classical
communications (ELOCC) convertibility decided by the R\'{e}nyi entropy \cite{key-17} suddenly changes
nearly at the critical point. In Ref.~\cite{key-18}, the local operations and classical
communications (LOCC) convertibility decided by the majorization relation \cite{maj} has also been investigated
in the one-dimensional transverse field Ising model \cite{ising}.
These significant results suggest that not only are the methods of QIP useful as alternative
signatures of quantum phase transitions, but also the study of quantum phase transitions can
offer interesting insight into QIP.

In this paper, we study the relationship between two kinds of ground-state local convertibility
and quantum phase transitions in XY model. The majorization relation studied here  is necessary and
sufficient for the LOCC convertibility \cite{maj} and R\'{e}nyi entropy interception is a necessary and
sufficient condition for the ELOCC convertibility \cite{key-19}. In the phase diagram of XY model,
LOCC convertibility and ELOCC convertibility of ground-states are presented and compared.
Since the one-dimensional transverse field Ising model can be seen as a special case of our model,
we promote and widen the results in previous literature. It is shown that different phases in the
phase diagram of XY model can have different LOCC or ELOCC convertibility  and both of them
changes around the phase transition point, which have successfully help us understand the deep
nature of the different phases in XY model. On the contrary, it will also benefit many applications
such as to detect the quantum phase transition and to estimate the computational power of
different phases in quantum critical systems \cite{key-16}.


\section{Criterions for LOCC and ELOCC convertibility}\label{sec:2}
\begin{figure}[t]
 \centering
\includegraphics[width=0.48\textwidth]{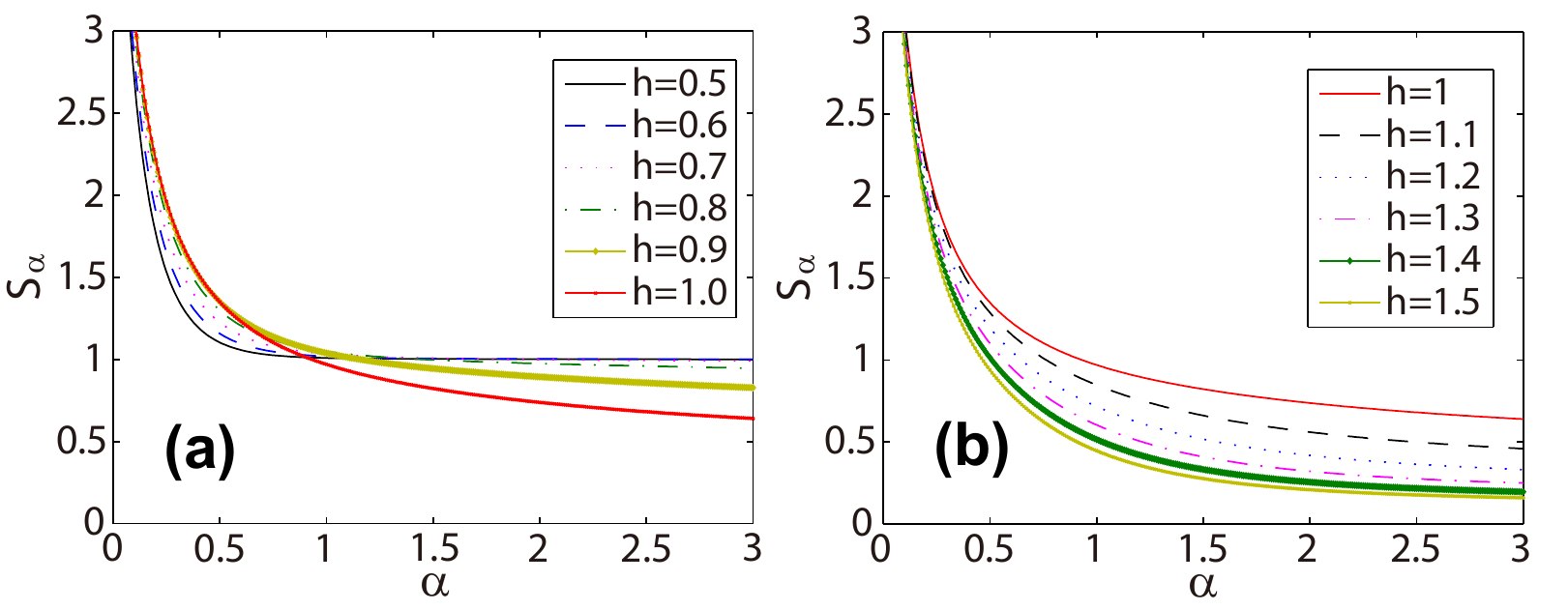}\\
\caption{(color online). The entanglement R\'{e}nyi entropy for the ground
state of the transverse field Ising model versus $\alpha$. The
R\'{e}nyi entropies are intercepted as (a) $h\leq 1$, while they
are non-intercepted as (b) $h\geq 1$, which gives a physical
significance to QIP from quantum transitions.}\label{fig:1}
\end{figure}

As in Ref.~\cite{key-16}, we consider a system with an adjustable external parameter $h$, which
partitioned into two parties, Alice and Bob, and operated by ELOCC or LOCC. Let $|G(g)\rangle_{AB}$ be
the ground state of the system when the parameter is set as $h$. Given the infinitesimal $\Delta$, the
necessary and sufficient condition for ELOCC between $|G(h)\rangle_{AB}$ and $|G(h+\Delta)\rangle_{AB}$
is given by the inequality $S_{\alpha}(\rho_{A}(h))\geq S_{\alpha}(\rho_{A}(h+\Delta))$ for all values of
$\alpha$, where $S_{\alpha}(\rho_{A})$ is the R\'{e}nyi entropy of the reduced state of Alice
$\rho_{A}=\textrm{Tr}_{B}(|G(h)\rangle_{AB}\langle G(h)|)$ (called the entanglement R\'{e}nyi
entropy of $|G(h)\rangle_{AB}$) defined as \cite{key-17}
\begin{eqnarray}
S_{\alpha}(\rho_{A}(h))=\frac{1}{1-\alpha}\log_{2}[\textrm{Tr}\rho_{A}(h)^{\alpha}].
\end{eqnarray}
When $\alpha\rightarrow1$, the R\'{e}nyi entropy tends to the von Neumann entropy.
In brief, there are two different behaviors for entanglement R\'{e}nyi entropies of
two states neighboring states $|G(h)\rangle_{AB}$ and $|G(h+\Delta)\rangle_{AB}$.
The transverse field Ising model with critical point $h=1$ is illustrated as an example
in Fig.~\ref{fig:1}: ($a$) If entanglement R\'{e}nyi entropies are crossing, those
two states can not be locally transferred to each other by ELOCC; ($b$) If there is no
crossing, a state with higher entanglement can be locally transferred to the lower
entanglement one by ELOCC  \cite{key-15}. These results can be applied to study the quantum
critical phenomena, i.e. when a quantum phase transition occurs, the local transformation
property of the ground state wave function changes as well as the different quantum
phases boundaries can be determined by the behavior of entanglement R\'{e}nyi entropies
of ground states \cite{key-15,key-16}.  Referring the scaling analysis in Ref.~\cite{key-15}
phase transition point obtained by this proposal tends to $0.9940$.

On the other hand, the majorization relations provide a necessary and sufficient condition
for the LOCC convertibility. Using the Schmidt decomposition theorem \cite{nc}, the
ground state can be written as
$|G(h)\rangle_{AB}=\sum_{k=1}^{d}\sqrt{\lambda_{h}^{k}}| k\rangle_{A}|k\rangle_{B}$.
Then, we have that
if and only if the majorization relations
\begin{eqnarray}
f_l(h+\Delta)\equiv\sum_{k=1}^{l}\lambda_{h+\Delta}^{k}\geq\sum_{k=1}^{l}\lambda_{h}^{k}\equiv f_l(h)
\label{eq:2}
\end{eqnarray}
are satisfied for all $1\leq l\leq d$ (expressed as $\lambda_{h+\Delta}\succ\lambda_{h}$,
state $|G(h+\Delta)\rangle$ can be transformed with 100\% probability of
success to $|G(h)\rangle$ by LOCC. Otherwise, these two states are incomparable,
i.e., $|G(h+\Delta)\rangle$ cannot be converted to $|G(h)\rangle$ by LOCC,
and vice versa. In Ref.~\cite{key-18}, it provides a clear description of the local
convertibility of the ground states of the transverse field Ising model:
in the region $0<h<0.9940$ within the ferromagnetic phase, neither
LOCC nor ELOCC convertibility exist; in the region $h>1.1318$
within the paramagnetic phase, both LOCC and ELOCC convertibility
exist; and in the small interval $0.994<h<1.1318$ around the critical
point $h=1$, merely the ELOCC convertibility exists.


\begin{figure*}[t]
 \centering
 \subfigure[]{\includegraphics[width=0.245\textwidth]{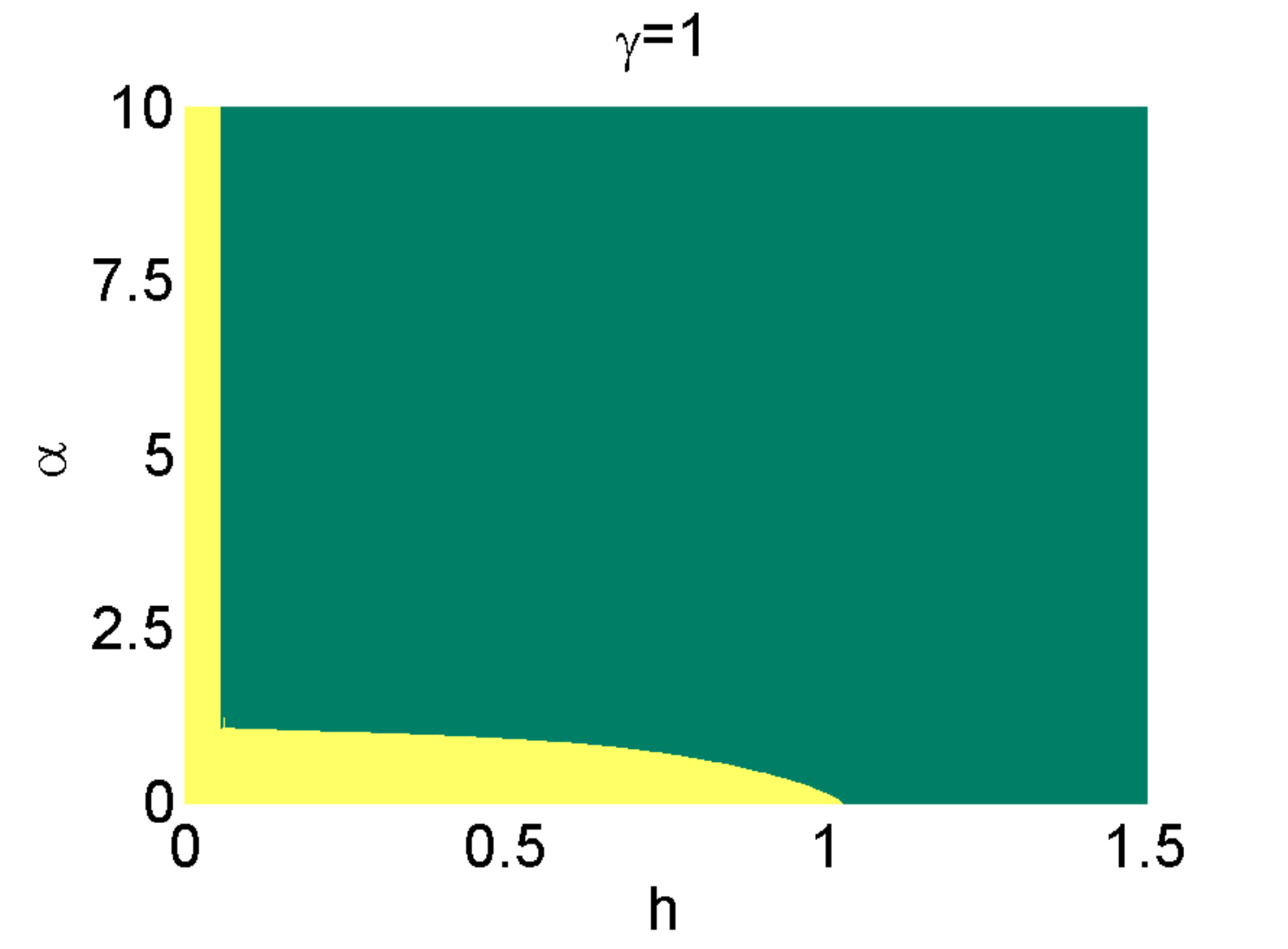}}
\subfigure[]{\includegraphics[width=0.245\textwidth]{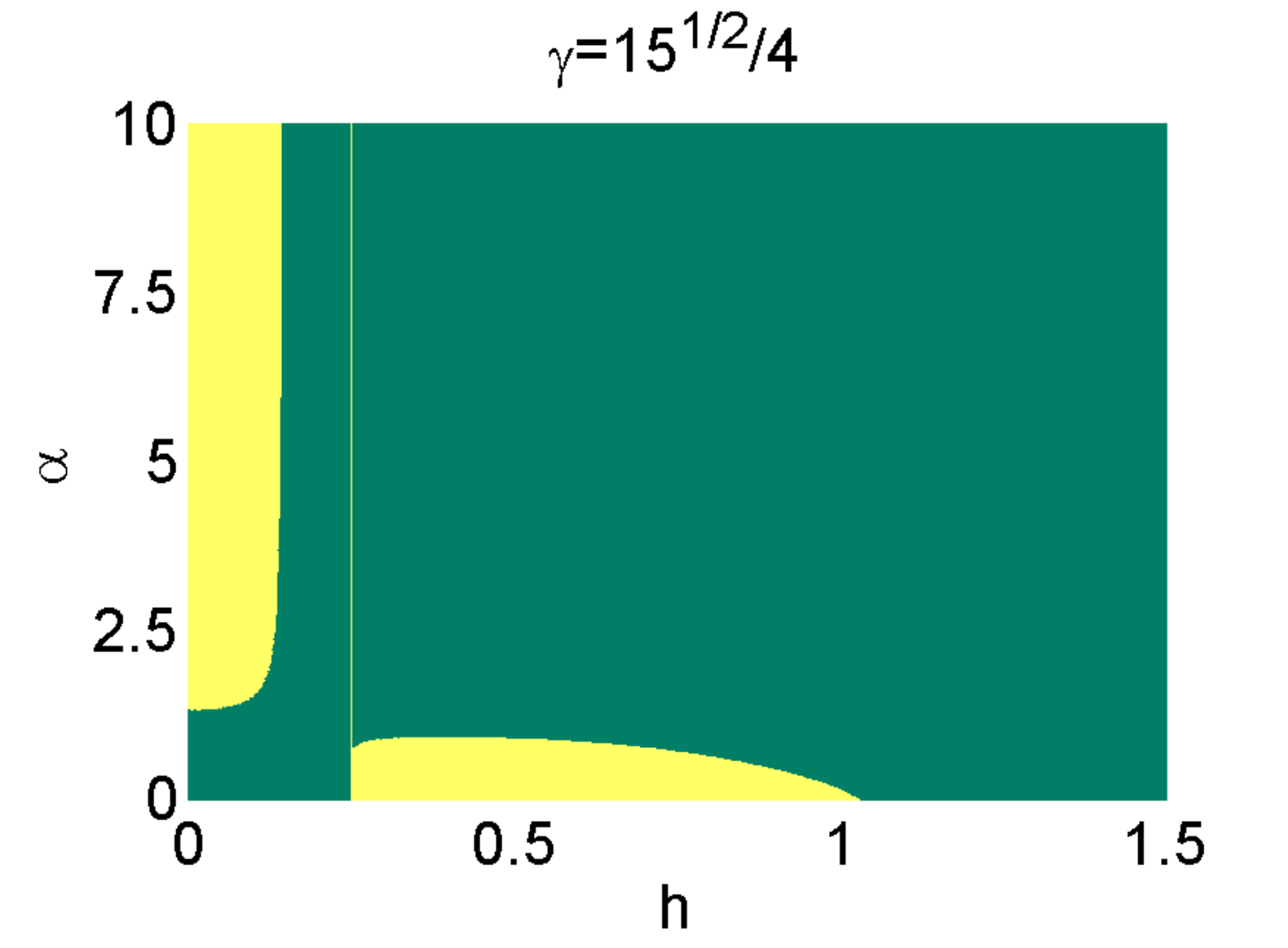}}
\subfigure[]{\includegraphics[width=0.245\textwidth]{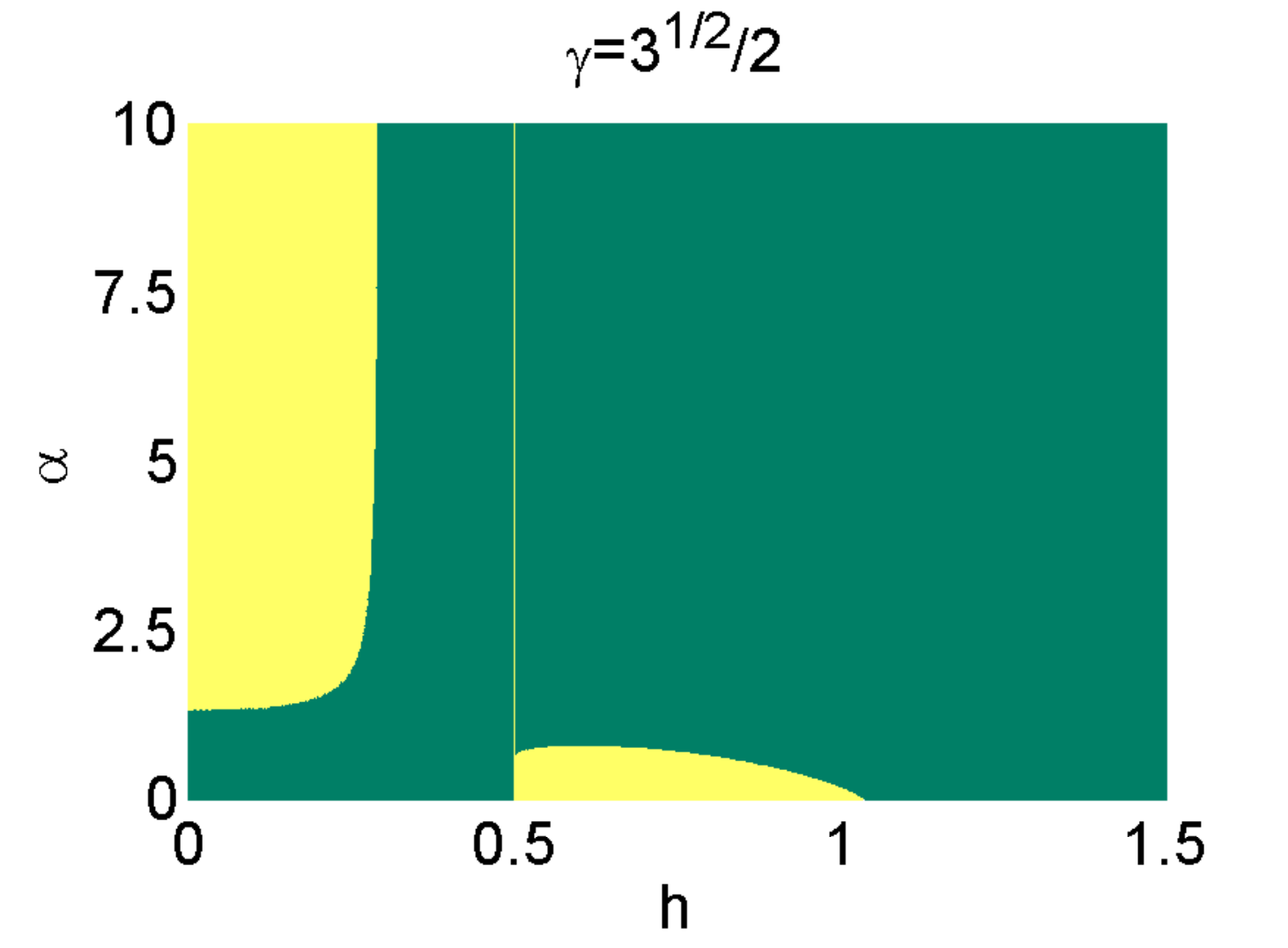}}
\subfigure[]{\includegraphics[width=0.245\textwidth]{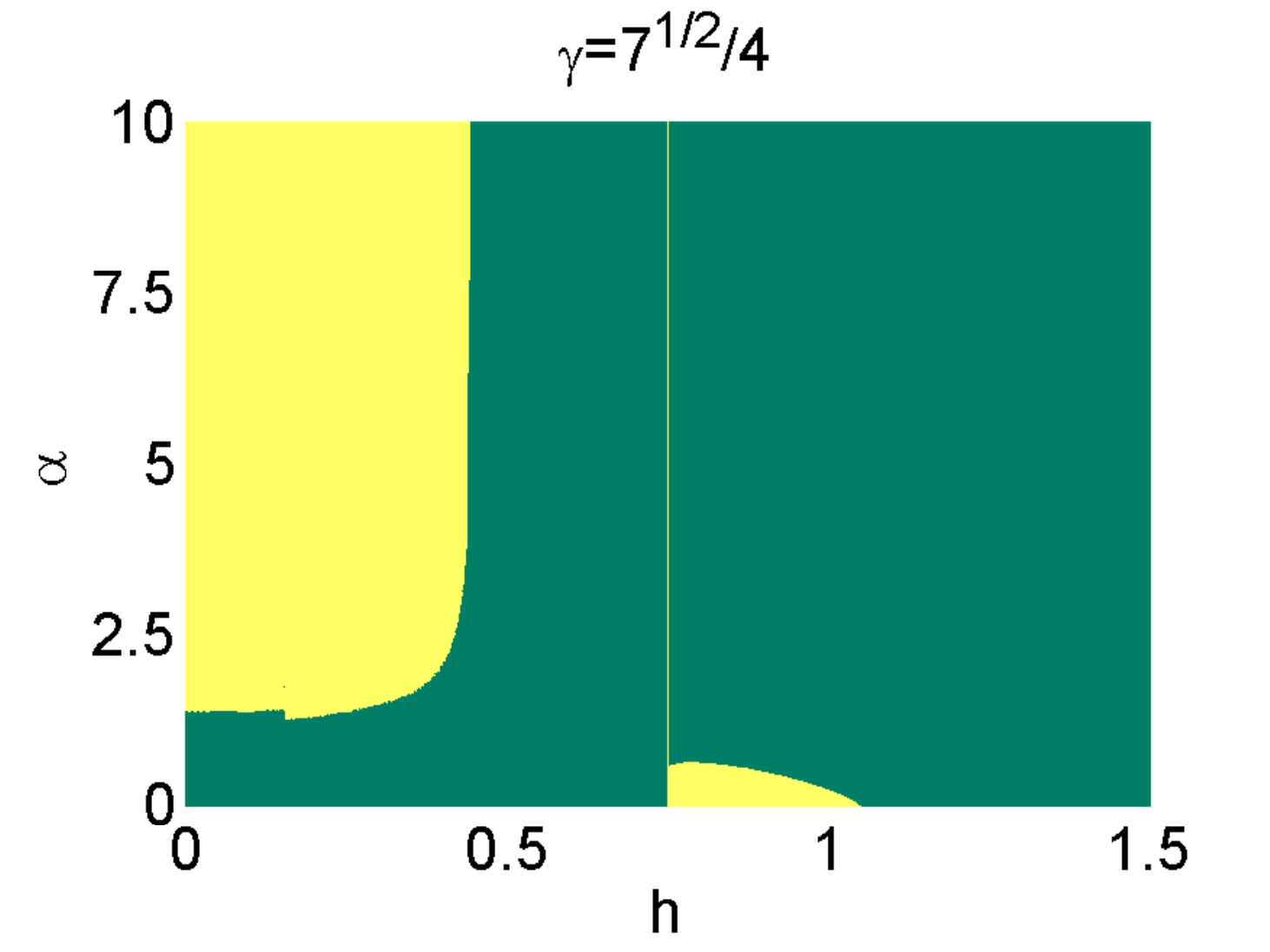}}
\subfigure[]{\includegraphics[width=0.245\textwidth]{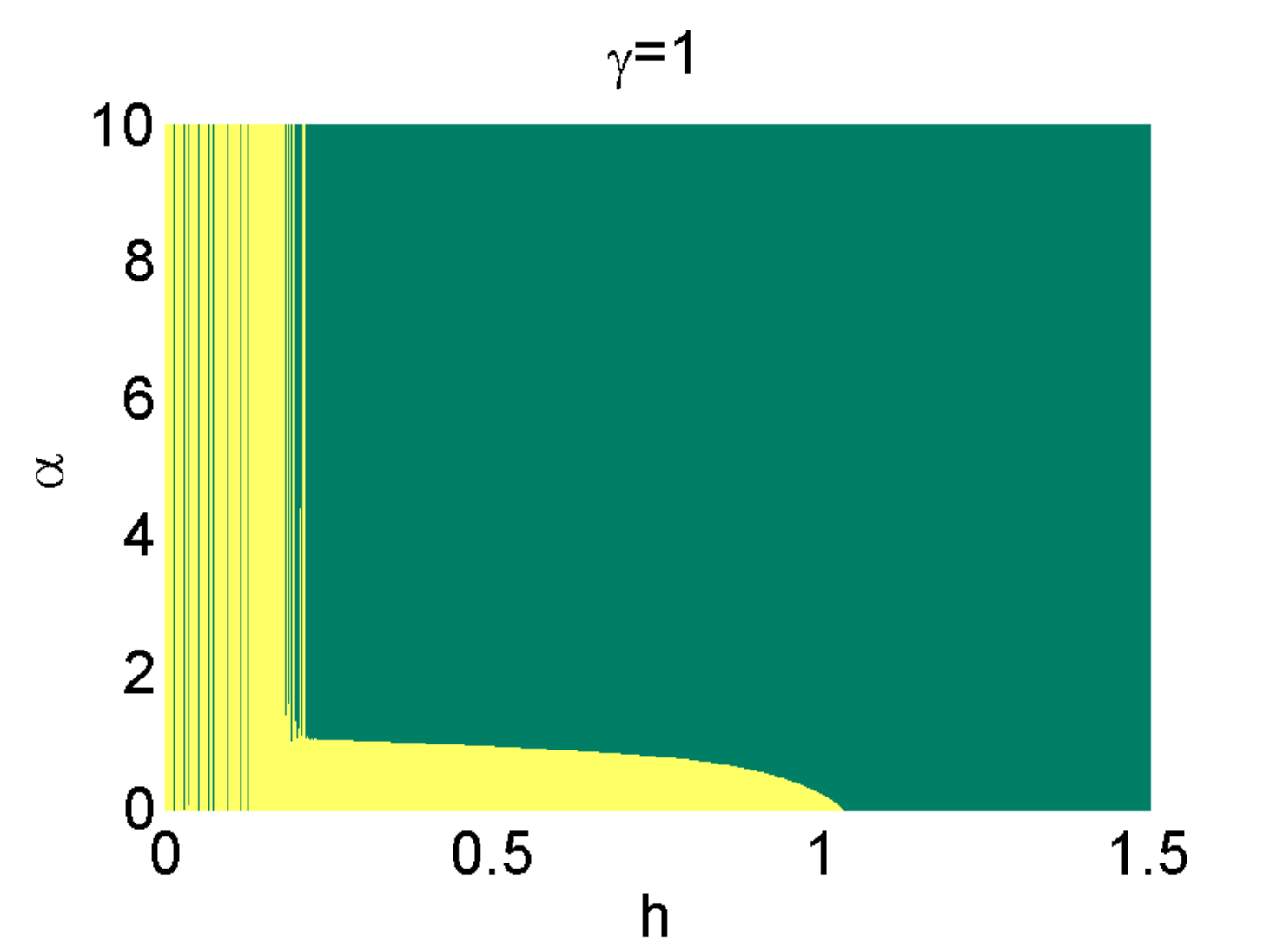}\label{fig:2e}}
\subfigure[]{\includegraphics[width=0.245\textwidth]{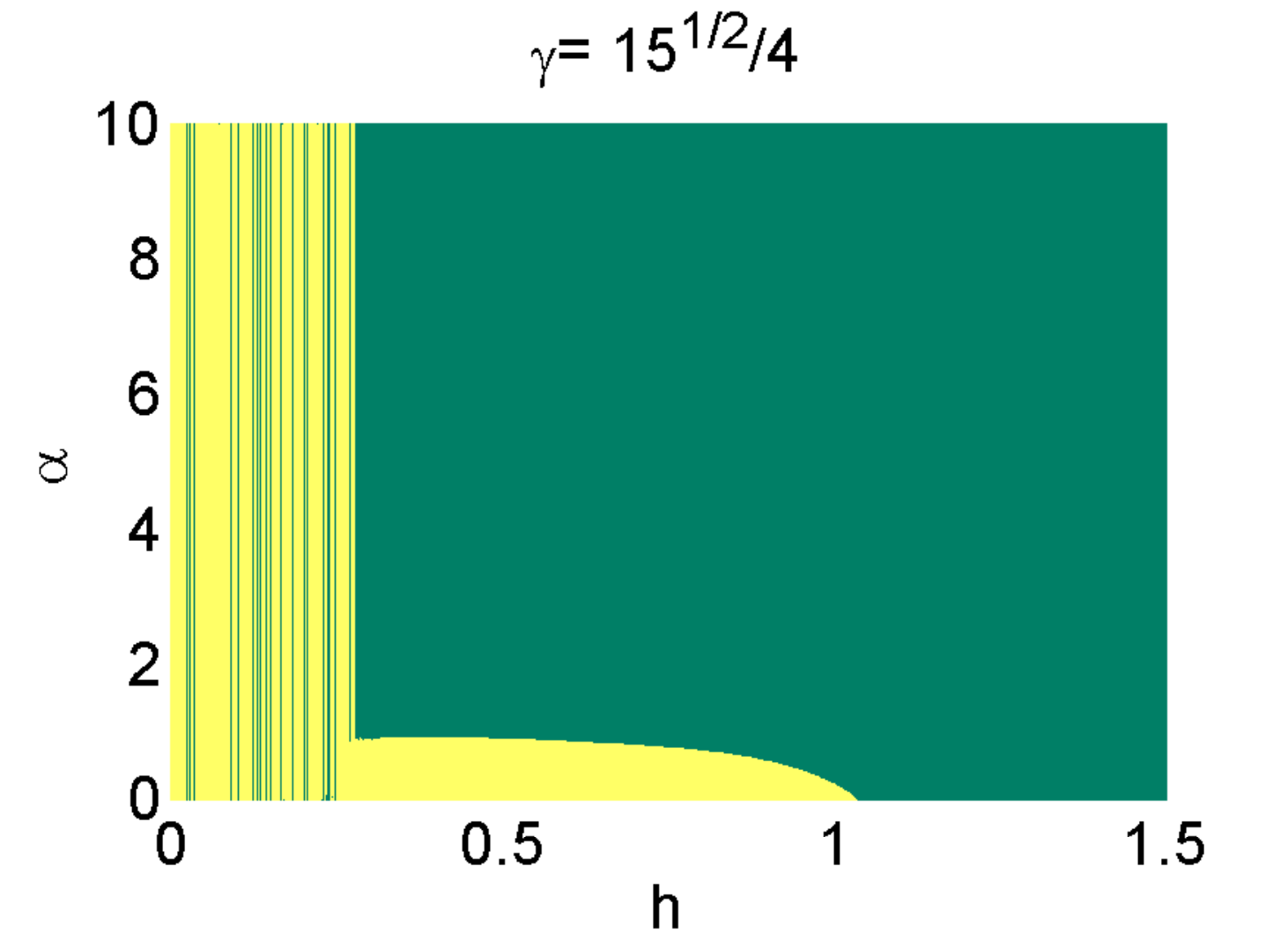}\label{fig:2f}}
\subfigure[]{\includegraphics[width=0.245\textwidth]{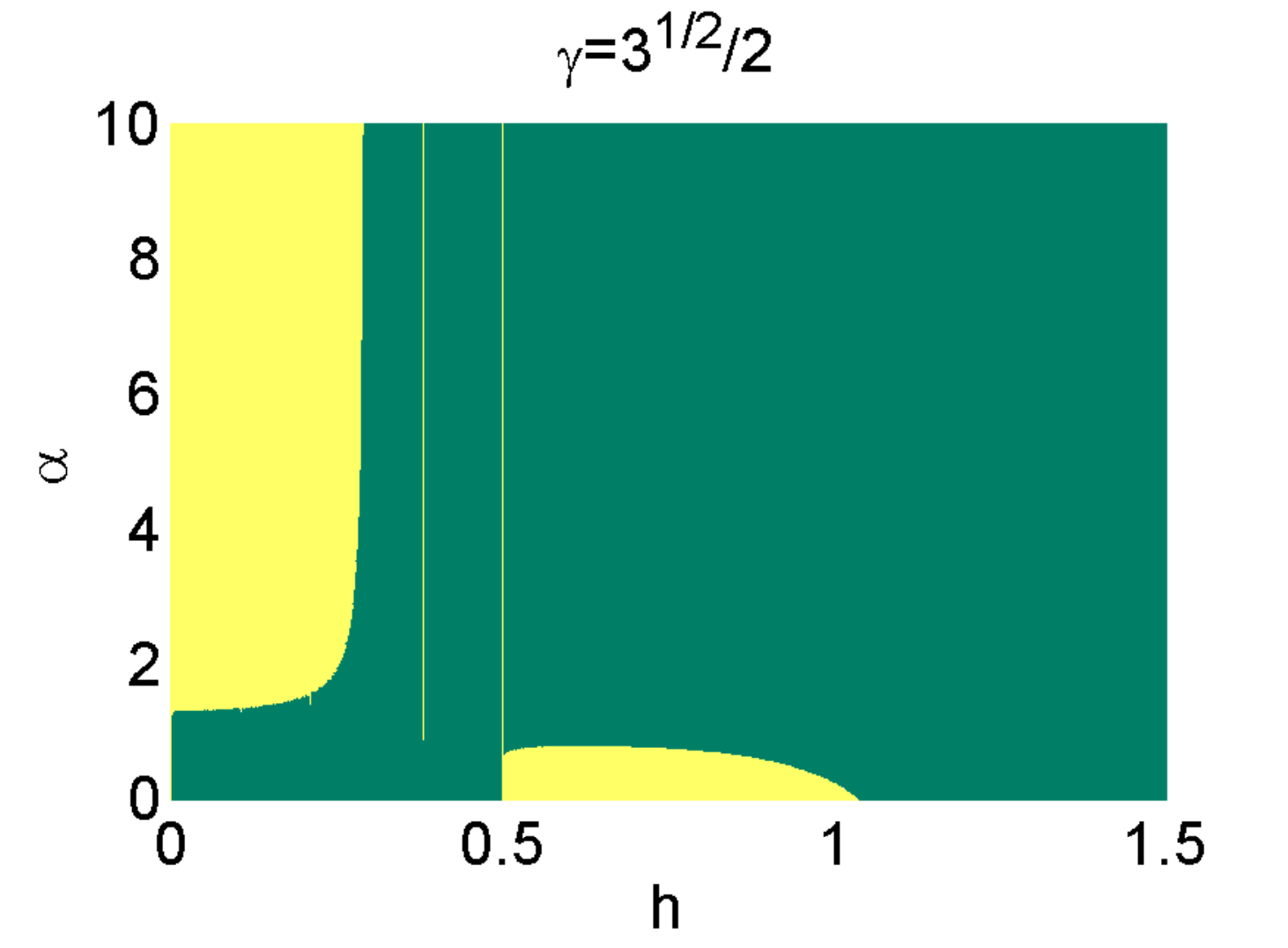}}
\subfigure[]{\includegraphics[width=0.245\textwidth]{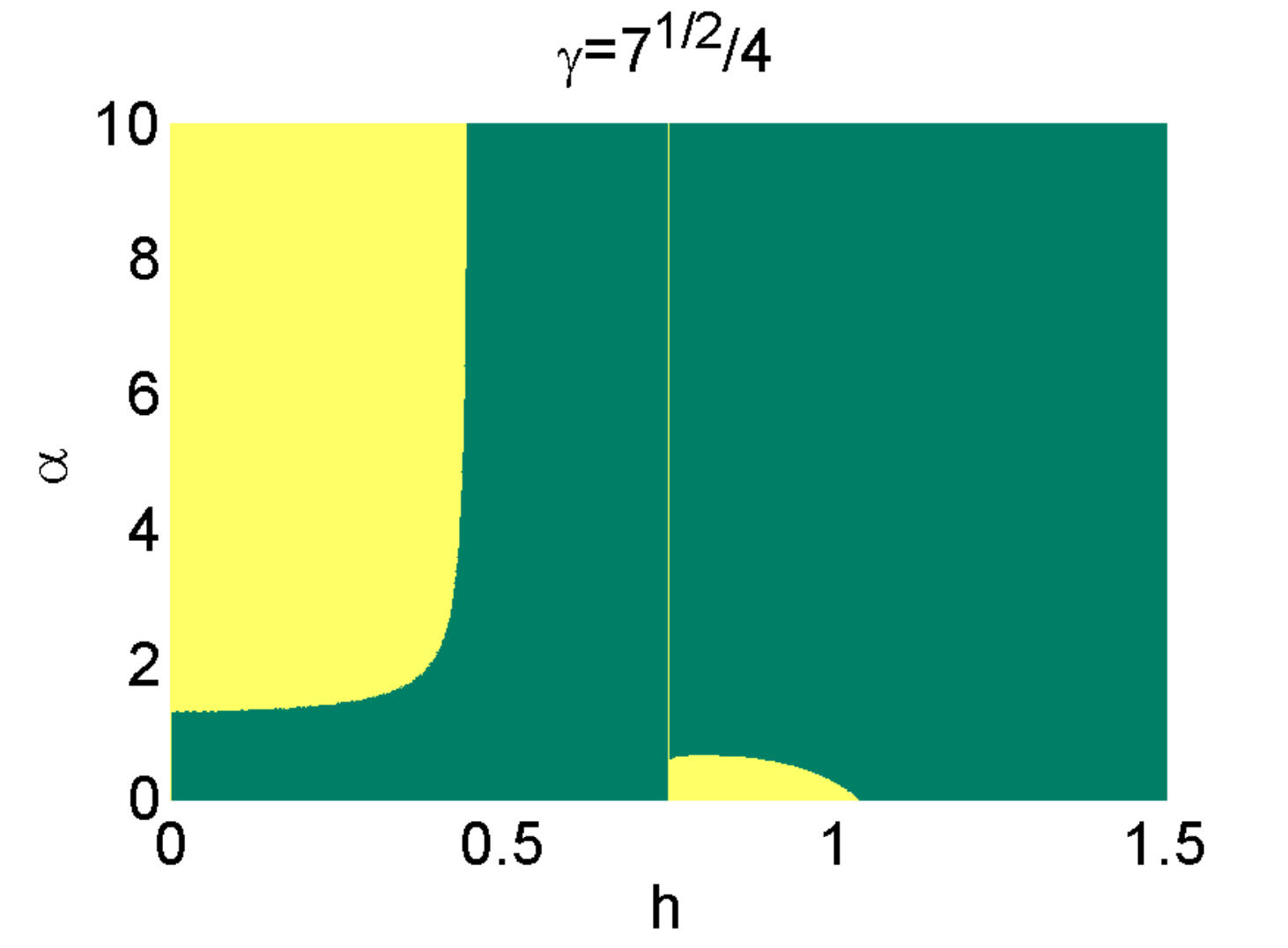}}
\\
\caption{(color online).  The sign distribution of $\partial_{h}S_{\alpha}(\rho_{A}(h))$ with
Alice having a $2$-spin system in XY model for different system size $L$ and parameter
$\gamma$ against $\alpha$ and the adjustable external parameter $h$. The green (dark) areas
represent the sign to be negative and the yellow (light) areas indicate positive. In (a-d) the system
size is $L=8$, and (e-h) show the case that the system size is $L=15$.  Note that the transition
between Phase 1A and 2 occurs at $h=1$, and the transition between Phase 1A and 1B occurs at
$h= 0.25$ for $\gamma = \sqrt{15}/4$ [see (b) and (f)], $h= 0.5$ for $\gamma =\sqrt{3}/2$ [see (c) and (g)], and
$h= 0.75$ for $\gamma =\sqrt{7}/4$ [see (d) and (h)].}\label{fig:2}
\end{figure*}

\section{Local convertibility of ground states of XY model}

Next, given the XY model \cite{key-20} in the zero-temperature case,
we use majorization relation and R\'{e}nyi entropy interception to study
the relationship between two kinds of local convertibility and quantum
phase transitions. The Hamiltonian of our model is as follows \cite{key-21}:
\begin{equation}
H=-\sum_{i=0}^{L-1}\left[\frac{1}{2}(1+\gamma)\hat{\sigma}_{i}^{x}\hat{\sigma}_{i+1}^{x}+\frac{1}{2}(1-\gamma)\hat{\sigma}_{i}^{y}\hat{\sigma}_{i+1}^{y}+h\hat{\sigma}_{i}^{z}\right]
\end{equation}
with $L$ being the number of spins in the chain, $\hat{\sigma}_{i}^{m}$
the $i$th spin Pauli operator in the direction $m=x,y,z$ and periodic
boundary conditions assumed. The XX model and transverse field Ising
model thus correspond to the special cases for this general class
of models. For the case that $\gamma\rightarrow0$, our model reduces
to XX model. When $\gamma=1$, the model reduces to transverse field
Ising model \cite{ising}.
For $h=1$, a second-order quantum phase transition takes place for any $0\leq\gamma\leq1$.
In fact, there exists additional structure of interest in phase space
beyond the breaking of phase flip symmetry at $h=1$. It\textquoteright{}s
worth noting that there exists a circle boundary between Phase 1A and Phase 1B, $h^{2}+\gamma^{2}=1$, on
which the ground state is fully separable. According to the previous
literature, this circle can be seen as a boundary between two differing
phases which are characterized by the presence and absence of parallel
entanglement \cite{key-22,key-23,key-24,key-25}. In fact, for each
fixed $\gamma$, the system is only locally non-convertible when
$h^{2}+\gamma^{2}>1$. Now we can divide the system into three separate
phases, Phase 1A, Phase 1B and Phase 2, where the ferromagnetic region
is now divided into two phases defined by their differential local
convertibility. These results are summarized in ``phase-diagram''
Fig.~\ref{fig:3a}.

We study the ground states of this model for different system sizes
 with the field parameter $h$ varying from $0$
to $1.5$. The ground states labeled as $| G\left(h\right)\rangle_{AB}$
are obtained by exactly diagonalizing the whole Hamiltonian $\left(3\right)$.
This proposal is also worth further investigations by other numerical
methods such as the Lanczos algorithm and density matrix renormalization
group (DMRG), which are not included in this work.

\subsection{ELOCC convertibility via entanglement R\'{e}nyi entropy}
Using this method mentioned in Sec.~\ref{sec:2}, we provide a phase diagram which describes the
ELOCC convertibility for different phases in XY model. We consider the case that Alice shared a
reduced system of two spins.

In Fig.~\ref{fig:3b}, for a system size $L=15$, we have two different
regions that indicate different ELOCC convertibility. In the yellow (light) areas, the sign of
$\partial_{h}S_{\alpha}(\rho_{A}(h))$ is negative for all $\alpha$, that is, the two states $| G(h)\rangle_{AB}$
and $| G(h+\Delta)\rangle_{AB}$ can be converted to each other by ELOCC.  In the red (dark) areas,
the sign of $\partial_{h}S_{\alpha}(\rho_{A}(h))$ is positive for all $\alpha$, there is no ELOCC
convertibility in these regions. Similar as Fig.~\ref{fig:2e} and \ref{fig:2f}, the obscure areas in
Fig.~\ref{fig:3b} reflect the level-crossings that redefine the ground state of the system
(which are evident from the spectrum of the model). Comparing Fig.~\ref{fig:3a} and
Fig.~\ref{fig:3b}, it is obvious that some boundaries between yellow (light) area and red (dark)
area can be used to detect the critical line for different phases. Specifically, in Phase 1A, the
ELOCC convertibility does not exist; in some area of Phase 1B and  the entire area of Phase 2,
the ground states $| G\left(h\right)\rangle_{AB}$ and $| G\left(h+\Delta\right)\rangle_{AB}$ can
be converted to each other by ELOCC. Therefore, the phase transition between Phase 1A and
Phase 2  and the one between Phase 1A and Phase 1B can be described precisely by the ELOCC
convertibility via entanglement R\'{e}nyi entropy.

\subsection{LOCC convertibility via majorization relation}

\begin{figure*}[t]
 \centering
\subfigure[]{\includegraphics[width=0.3\textwidth]{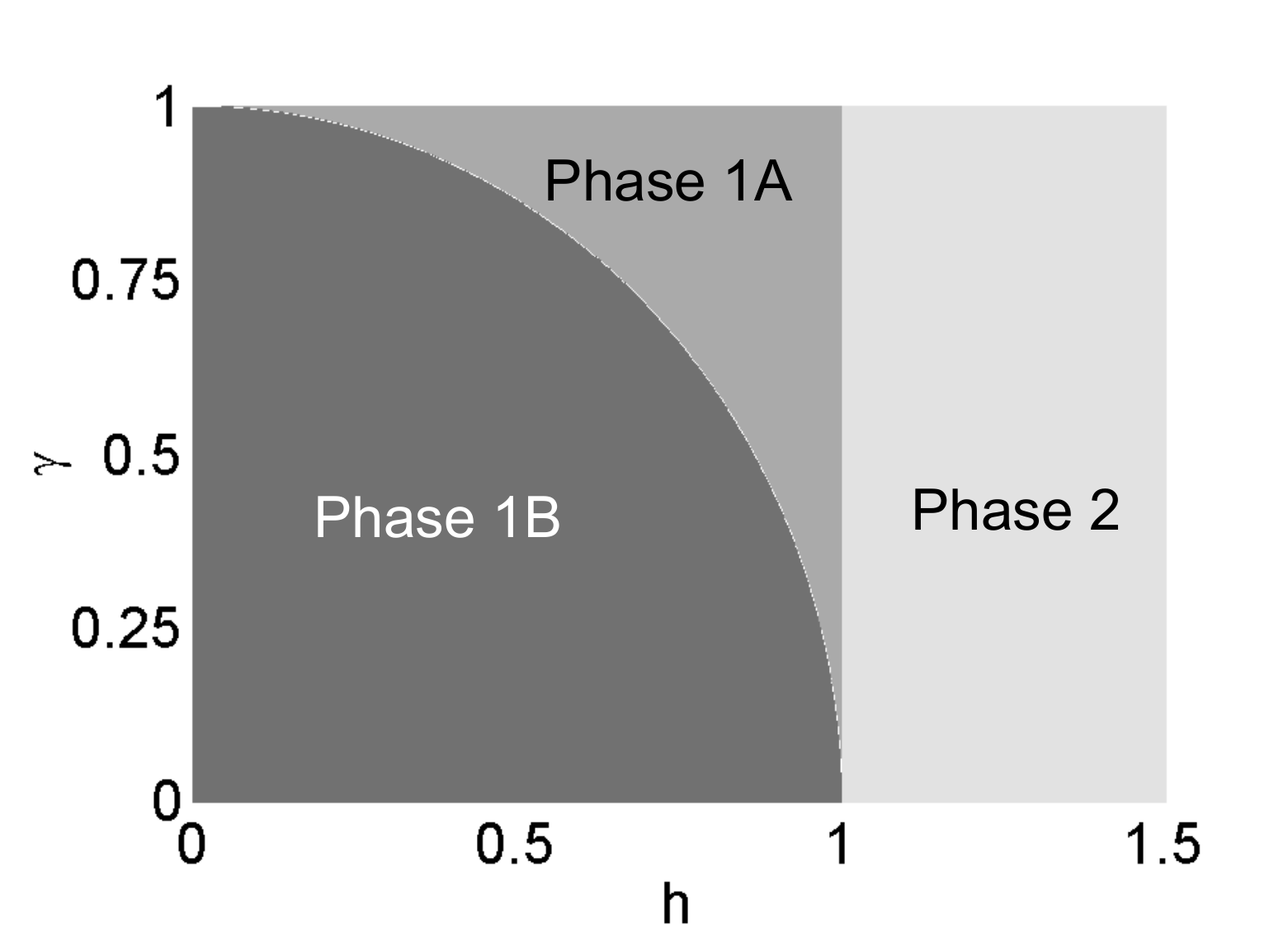}\label{fig:3a}}
\subfigure[]{\includegraphics[width=0.3\textwidth]{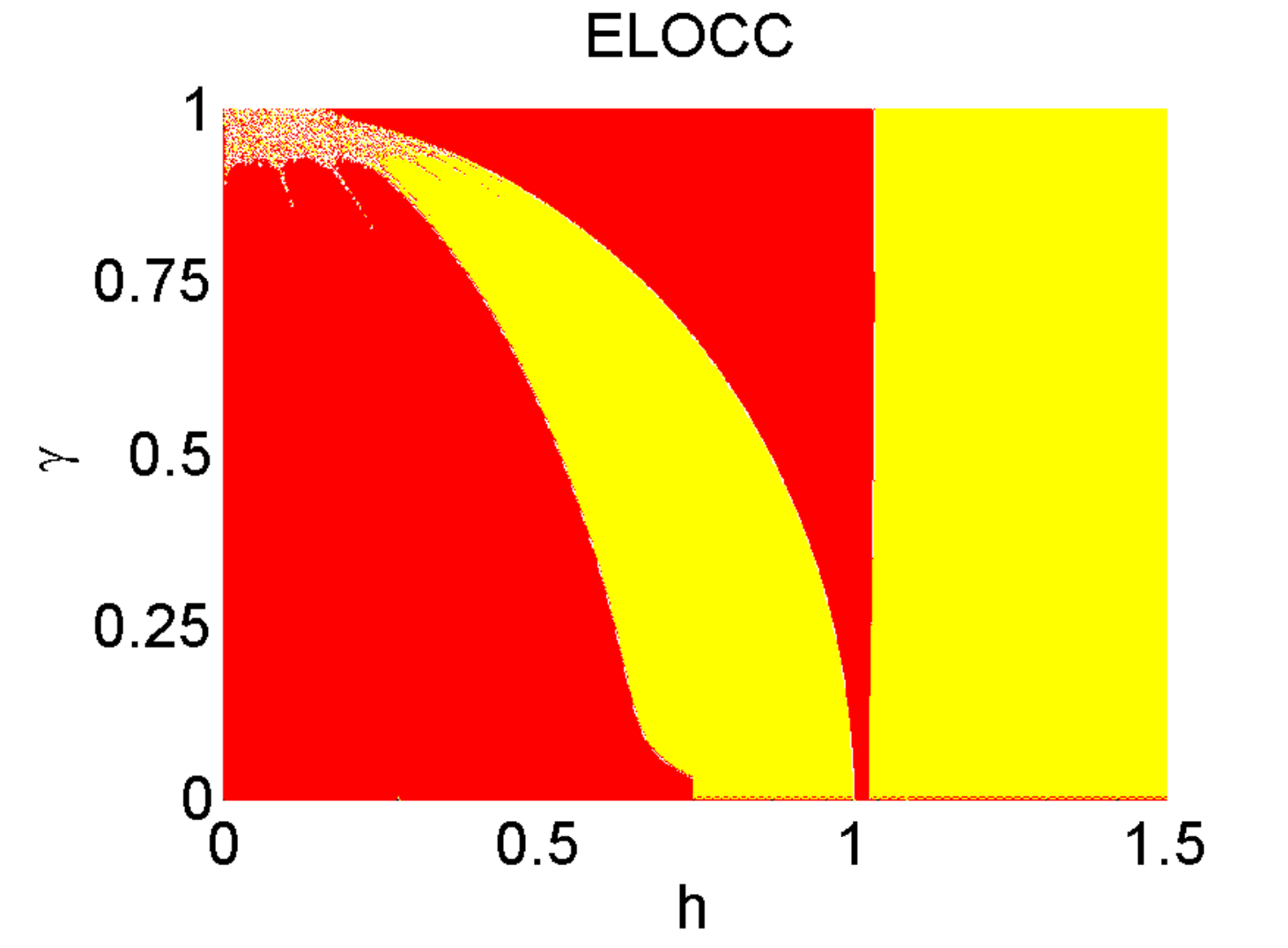}\label{fig:3b}}
\subfigure[]{\includegraphics[width=0.3\textwidth]{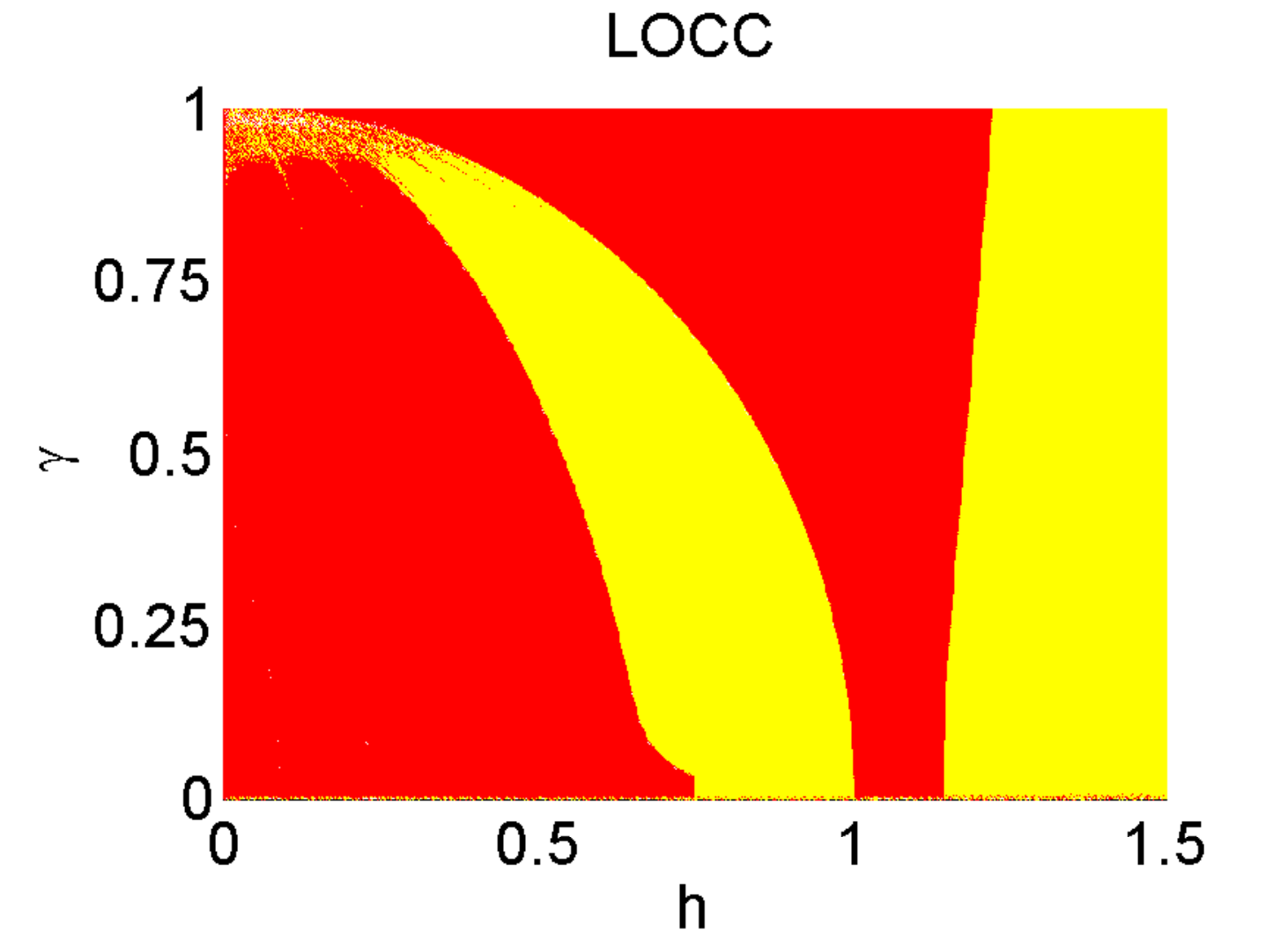}\label{fig:3c}}\\
\caption{(color online). (a) Phase diagram of XY model. XY model has three different quantum
phases: Phase 1A, Phase 1B and Phase 2. (b) ELOCC convertibility via R\'{e}nyi entropy of ground
states of XY model. In yellow (light) area, the sign of $\partial_{h}S_{\alpha}(\rho_{A}(h))$ is negative for all $\alpha$
which indicates that $| G\left(h+\Delta\right)\rangle_{AB}$ can be
transformed to $| G\left(h\right)\rangle_{AB}$ by ELOCC with 100\% probability
of success. The red (dark) areas are for other cases.  (c) LOCC convertibility via majorization of ground
states of XY model. In yellow (light) area, $| G\left(h+\Delta\right)\rangle_{AB}$ can be
transformed to $| G\left(h\right)\rangle_{AB}$ by LOCC with 100\% probability
of success; whilst, in the red (dark) area, $| G\left(h+\Delta\right)\rangle_{AB}$ can not be
transformed to $| G\left(h\right)\rangle_{AB}$ by LOCC with certainty.}\label{fig:3}
\end{figure*}

For the same case that the reduced system shared by Alice has two spins, we only need to consider
three largest eigenvalues of the reduced states of any
two neighboring spins as $\lambda_{1}\left(h\right)$, $\lambda_{2}\left(h\right)$,
and $\lambda_{3}\left(h\right)$ in descending order. To detect the
majorization relations between two ground states $| G\left(h\right)\rangle_{AB}$
and $| G\left(h+\Delta\right)\rangle_{AB}$ given some infinitesimal
$\Delta$, we should judge the monotonicities of three functions
for the field parameter $f_1(h)$, $f_{2}(h)$ and $f_{3}(h)$ defined in Eq.~(\ref{eq:2}).
Thus, three cases will be met: ($i$) when monotonicities of these
three functions are all non-increasing, $| G\left(h+\Delta\right)\rangle_{AB}$
can be converted to $| G\left(h\right)\rangle_{AB}$ by LOCC with 100\% certainty;
($ii$) when monotonicities of these three functions all are non-decreasing,
$| G\left(h\right)\rangle_{AB}$ can be converted to $| G\left(h+\Delta\right)\rangle_{AB}$
by LOCC with 100\% certainty; ($iii$) except for these two cases, no state
can be converted to the other via LOCC with certainty. For ease of comparison with ELOCC
convertibility, We only show in Fig.~\ref{fig:3c} either the case ($ii$) or other two cases.

Using the method mentioned above, we can also provide a phase diagram describing
the LOCC convertibility for different phase in XY model. In Fig.~\ref{fig:3c},
for system sizes $N=15$, we have shown two different regions. 
In the yellow (light) area, $| G\left(h+\Delta\right)\rangle_{AB}$ can be
transformed to $| G\left(h\right)\rangle_{AB}$ by LOCC with 100\% probability
of success; whilst, in the red (dark) area, $| G\left(h+\Delta\right)\rangle_{AB}$ can not be
transformed to $| G\left(h\right)\rangle_{AB}$ by LOCC with certainty. We should stress that in few
parts of red (dark) area in Phase 1A, $| G\left(h\right)\rangle_{AB}$ can be transformed to
$| G\left(h+\Delta\right)\rangle_{AB}$ by LOCC with certainty, which is not shown in
Fig.~\ref{fig:3c}.
%
It is worth noting that in the region $g<0.5$, we can see an obscure area, which reflects
the level-crossings redefining the ground state of the system. In the XY model, there are three phases,
Phase 1A, Phase 1B and Phase 2 as shown in Fig.~\ref{fig:3a}. From Fig.~\ref{fig:3c}, in Phase 1A, no majorization
relations can be fulfilled, 
$| G\left(h+\Delta\right)\rangle_{AB}$ and $| G\left(h\right)\rangle_{AB}$ can not be converted to each other
by LOCC. However, in Phase 1B and Phase 2, there are two areas which have different
LOCC convertibility, which have richer and more complex features of LOCC convertibility than those of ELOCC.
Similarly, we can use the dividing line between red (dark) and yellow (light) areas with different LOCC convertibility
to detect the critical line between Phase 1A and Phase 1B precisely.
As the increase of the system size $L$, our result can be shown to get better using the
scaling analysis on the critical point from LOCC convertibility \cite{key-18} and ELOCC convertibility \cite{key-15}
to get rid of the finite-size effect.

Comparing the result with the ELOCC method, we can see that the ELOCC method is superior
to LOCC metohd for detecting the critical line between two different quantum phases.
In order to detect the phase transitions in our model, the ELOCC method is more
accurate than LOCC method in the XY model as well as in the one-dimensional transverse
field Ising model \cite{key-18}. 
Moreover, we should note that if the LOCC convertibility exists, the ELOCC convertibility must exist.
That is, in different phases of quantum critical systems, various and complicated local conversion
can be expected, will benefit many applications such as to estimate the computational power of
different phases.

\begin{figure}[b]
 \centering
\includegraphics[width=0.35\textwidth]{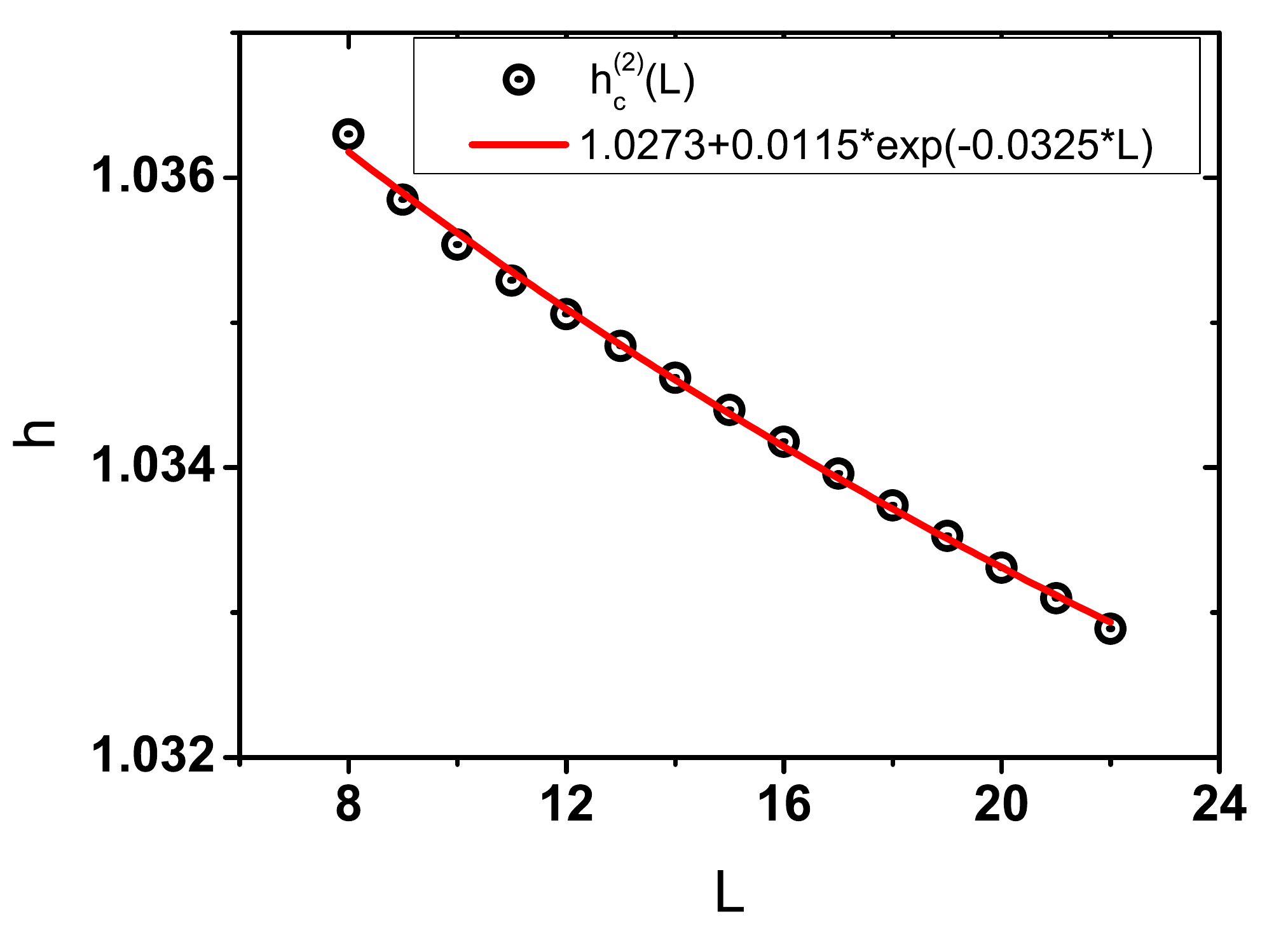}\\
\caption{(color online). Finite-size scaling analysis of the the critical points $h^{(1)}_c(L)$ of the second-order
phase transition of XY model by the ELOCC convertibility
for $\gamma=\sqrt{3}/2$.}\label{fig:4}
\end{figure}

To get rid of the finite-size effect, we give a scaling analysis of
the critical points $h^{(2)}_{c}(L)$ of the second-order phase transition detected by the ELOCC convertibility
for $\gamma=\sqrt{3}/2$ in Fig.~\ref{fig:4}. With the exponential fitting, it is shown that for infinitely large
size $L\rightarrow\infty$ and $\gamma=\sqrt{3}/2$, the ELOCC convertibility changes at $h^{(2)}_c=1.0273$,
which detects the second-order phase transition of the XY model. For the first-order phase transition of the
XY model, the ELOCC convertibility changes at exactly $h_c^{(1)}=0.5$ for all sizes. The scaling analysis of critical
points detected by the LOCC convertibility can be found in Ref.~\cite{key-18}.

\section{Conclusion}
In conclusion, we have investigated the relationship between local convertibility of ground states and quantum phase transitions in XY model.
We study the LOCC convertibility by examining the majorization relations
and the ELOCC convertibility via R\'{e}nyi entropy interception. It shows
that the boundary of areas which have different LOCC or ELOCC convertibility
can be used to detect the critical line for different phases. In
Phase 1A, both the LOCC and ELOCC convertibility do not exist. In
Phase 1B, the situation is slightly more complicated. There are two
areas which has different LOCC convertibility. In most area of Phase
1B, the ELOCC convertibility exists. In Phase 2, there is no LOCC
convertibility in the area that nearly the critical point. In the
remaining area, both the LOCC and ELOCC convertibility exist. It is
obvious that if the LOCC convertibility exists, the ELOCC convertibility
must exist. 
We believe that our method should be applicable in other systems
similar as in the XY model. The LOCC proposal in detecting quantum
phase transition can play a complementary role to the ELOCC method,
which will help to exploit the rich features of local conversion of ground states of quantum critical systems.
This study will enlighten extensive studies of quantum phase transitions
from the perspective of local convertibility, e.g., finite-temperature
phase transitions and other quantum many-body models.

The authors thank Jun-Peng Cao, Dong Wang, Yu Zeng, and Shuai Cui 
for their valuable discussions. This work was supported by the
NSFC (Grant Nos. 11175248, 11375141, 11425522, 11434013), the grant from Chinese Academy of Sciences (XDB01010000).

\end{document}